\documentclass{ws-I2}

\def\cc 
\def\tc{\Blue} 
\def\oc{} 

\def\bm#1{\mbox{\boldmath{$#1$}}}
\def\ra{\rightarrow}
\def\D{{\cal D}}
\def\P{{\cal P}}

\begin{document}

\title{The exact renormalization group in Astrophysics}

\author{Jos\'e Gaite}

\address{Centro de Astrobiolog\'{\i}a, 
CSIC-INTA,\\
Carretera Ajalvir, km 4, 
28850 Torrej\'on de Ardoz, Madrid, Spain
}


\maketitle

\abstracts{The coarse-graining operation in hydrodynamics is
equivalent to a change of scale which can be formalized as a
renormalization group transformation. In particular, its application
to the probability distribution of a self-gravitating fluid yields an
``exact renormalization group equation" of Fokker-Planck type.  Since
the time evolution of that distribution can also be described by a
Fokker-Planck equation, we propose a connection between both
equations, that is, a connection between scale and time evolution.  We
finally remark on the essentially non-perturbative nature of
astrophysical problems, which suggests that the exact renormalization
group is the adequate tool for them.}

\section{Dynamical equations and fluctuations}

\subsection{Astrophysical hydrodynamics}

The usual hydrodynamic equations are the continuity equation, the
Euler (or Navier-Stokes) equation describing momentum conservation,
and the energy equation.  To account for the gravitational field, we
must add its corresponding equations, namely, the usual equations for
its divergence and curl. In cosmology, we must also account for the
global expansion, which, in the Newtonian formulation, introduces the
scale factor $a(t)$ and the Hubble constant $H(t)$.\cite{Pee}

Then we have five equations altogether to describe the evolution of
the fields, but we assume that the solution of the energy equation is
a {\em polytropic} equation of state $P = \kappa \,\varrho^\gamma$. So
we have four equations:
\bea
  \frac{\partial \varrho}{\partial t} \oc{+ 3 H \varrho} + \oc{{1 \over a}} 
  \nabla \cdot (\varrho \, {\bm v}) = 0,\\
   \frac{\partial {\bm v}}{\partial t} \oc{+ H \, 
  {\bm v}} + \oc{{1 \over a}}\, {\bm v}\cdot \nabla{\bm v} = 
  {\bm g} - \oc{{1 \over a}}\, {\nabla P\over \varrho},\\
\nabla \cdot {\bm g} = -4\pi G\,\oc{a}\,(\varrho -
\oc{\varrho_b}),\quad \nabla \times {\bm g} = {\bm 0};
\eea
$\varrho_b$ is the background (average) density.

Of course, the two equations for the gravitational field can be
reduced to a single equation for the gravitational potential. Then we
have two scalar equations, that is, the continuity and the Poisson
equation, and one vector equation, namely, the Euler velocity
equation, for two scalar unknowns (the density and the
gravitational potential) and one vector unknown (the
velocity). $H$, $\varrho_b$ and $a$ are known functions of time in a
definite cosmological model. For astrophysical situations in which the
cosmological expansion is irrelevant, we may put $a=1$ and
$\varrho_b=0$, $H=0$. 

 
\subsection{Coarse graining}\label{}

The hydrodynamic equations are macroscopic equations, which can be
assumed to follow from microscopic Newtonian mechanics of particles
through an averaging process called {\em coarse graining}.
The average density $\varrho_L({\bm r})$ of an element of volume, say
a box of side $L$, centered on the point ${\bm r}$ is the total mass of
the particles inside divided by the volume $L^3$. One defines similarly
the velocity field ${\bm v}_L({\bm r})$, etc. 

We must regard that the coarse graining operation does not completely
eliminate the irregular nature of the microscopic variables but leaves
a fluctuating component, which should vanish as $L$ grows. In general,
the use of hydrodynamics implies a thermodynamic treatment of the
element of volume, to which one can add the usual thermodynamic
fluctuations. However, since the gravitational force is long ranged,
the separation of scales on which the theory of thermodynamic
fluctuations is based is useless here. Therefore, the hydrodynamic 
equations must be supplemented with the fluctuations, as 
usual in hydrodynamics, but with fluctuations that are of a particular nature.%
\footnote{A recent discussion of hydrodynamic equations in the
presence of gravitation is given by A. Dom\'{\i}nguez.\cite{Alv}}  As
one scales $L$ up, the coarse-grained fields change, as well as the
fluctuations.  This is the essence of the renormalization group.
      
Besides stochastic fluctuations, which appear as {\em noise} terms in
the hydrodynamic equations, one can also consider a
probability distribution of initial fields, as usually done in
cosmology. The fluctuations are determined by the correlation
functions of the noise or the initial conditions.

\subsection{Dynamical evolution}

To solve a set of stochastic partial differential equations (SPDE),
one must solve before the partial differential equations as if the
noise were just an external source, and then express the correlation
functions of the dynamical fields $\varrho$ and ${\bm v}$ in terms of
the correlation functions of the noise (or initial conditions). Of
course, the solution is of stochastic nature and can also be expressed
in terms of a probability distribution $\P[\varrho,{\bm v},t]$: the
correlation functions are given by the corresponding functional
integrals.  The SPDE are therefore equivalent to an evolution equation
for the probability distribution $\P$. We shall show below how to
solve the hydrodynamic SPDE and the associated equation for $\P$
corresponding to a simplified version of the total set of equations.

\section{The exact renormalization group equation}

Apart from their time evolution, the coarse-grained variables change
(we may also say ``evolve'') with the coarse-graining length $L$. 
>From a field-theory point of view, a smoothing of the ``bare'' fields 
$\varrho$ and ${\bm v}$ is necessary to {\em regularize} short-distance 
singularities in functional integrals. It is convenient 
in our context to define a general coarse graining procedure with 
the help of a {\em window function}, that is, a function that quickly 
vanishes outside a neighborhood of the origin of size $L$; 
typical examples are the sharp-cutoff (``top-hat'') window and 
the Gaussian window. Thus, we define, for example, a smoothed density field 
as the convolution
\be
\varrho_L({\bm r}) = \int W_L({\bm r} - {\bm x})\,\varrho({\bm x}).
\ee

As $L$ runs, the coarse-grained variables change and so does the
probability distribution $\P$ (for simplicity, here we disregard its
time dependence, as well as the variable ${\bm v}$, and write
$\P_L[\varrho]$):
\be
{\partial\over \partial L}\P_L[\varrho] = {d \ln{\tilde W}_L\over
dL}\P_L[\varrho] -{d \ln{\tilde W}_L^2\over
dL}\,{\delta \over \delta\varrho}(\varrho\,\P_L[\varrho])-
{1\over 2}\,{d {\tilde W}_L^2\over
dL}\,{\delta^2 \over \delta\varrho^2}\P_L[\varrho],\label{ERGeq}
\ee
where ${\tilde W}_L$ is the Fourier transform of the
window function. This differential equation, describing evolution with 
the scale $L$, is the {\em exact renormalization group} equation for 
$\P_L$. Its proof is exposed in the appendix. A useful remark is that 
it is a sort of Fokker-Planck equation for $L$-evolution.

The exact renormalization group equation relies on Wilson
non-perturbative philosophy of the renormalization group and, in
fact, was proposed by Wilson himself.\cite{WilKo} It has been amply
used in high-energy and statistical physics, appearing in various
forms, essentially equivalent.  However, while coarse graining is
widely used in astrophysical hydrodynamics and the effect of changing
the coarse-graining length has been discussed on several occasions,
the exact renormalization group equation does not seem to have
appeared before in the astrophysical literature.%
\footnote{Nevertheless, it has been noticed that the point probability
distribution $p_L(\varrho) := \P_L[\varrho({\bm r})]$ satisfies a
diffusion equation, analogous to Eq.~(\ref{ERGeq}), but with only the
last term of the right hand side.\cite{Bond} It was used to find the
distribution of collapsed objects with volume $L^3$ according to
their mass.}

\section{Dynamical renormalization group}

Dynamical scaling arguments, customary in the physics of surface
growth,\cite{Stan} lead one to the conclusion that, when the initial
condition is forgotten, the dynamical variables adopt a scaling form
such as, for example, $\delta\varrho_L := \varrho_L - 
\langle \varrho \rangle \sim L^\alpha f(t/L^z)$, where
$\lim_{x \ra\infty} f(x) =1$ and $\lim_{x \ra 0} f(x) \sim
x^{\alpha/z}$. In words, the fluctuations grow with time as a power law and
they eventually reach saturation, in which state they depend on the
coarse-graining length as another power law.  An elementary example is
the (linear) diffusion equation, in which $z=2$.

The problem is to derive the exponents $\alpha$ and $z$ and the scaling
function $f$. Here is where the dynamical RG comes into play:
similarly to the situation in static critical phenomena, in which
the possible types of critical behaviour correspond to fixed points of
the ordinary RG, the possible types of critical dynamics correspond to
fixed points of the dynamical RG.

This philosophy is appropriate for the problem on hand, as attested by
computer simulations of the microscopic Newtonian mechanics of
particles.  Indeed, one observes that the fluctuations grow with
time. Moreover, if for a fixed time one reduces the coarse-graining
length, one observes a similar growth of fluctuations. It is natural
to expect that the asymptotic state corresponds to a fixed point of
the dynamical RG. Relying on this philosophy, this RG has been
employed to determine the fixed points and critical exponents of a
simplified hydrodynamic self-gravitating model.\cite{all}

Unfortunately, the usual treatment of the dynamical RG is
perturbative. This may render it inapplicable for self-gravitating
models, as I will argue later. Besides, the fixed point structure
depends on the properties of the noise. Both problems are overcome by
the exact renormalization group as we have derived it: it is
non-perturbative and independent of the noise.  However, it may depend
on the window function.  We shall see that this dependence is
equivalent to a dependence on the noise and enlightens the role of the
latter in the dynamics.

\section{Burgers and Kardar-Parisi-Zhang equations}

In cosmology, it is customary to resort to the Zel'dovich (or
parallelism) approximation, which constrains the velocity to have the
same direction that the gravitational field has, that is, ${\bm
v}({\bm r},t) = F(t)\,{\bm g}({\bm r},t)$.\cite{Pee} This
approximation greatly simplifies the set of hydrodynamic equations, in
fact, reducing them to a single equation for the (redefined)
velocity. Furthermore, in the case that the polytropic index $\gamma =
2$, this equation becomes the three-dimensional Burgers equation,
giving rise to the {\em adhesion model}.\footnote{The one-dimensional
Burgers equation had been introduced earlier as a toy model for
turbulence.\cite{Burgers} Its 3D counterpart represents
pressure-less compressible turbulence without vorticity.}
If we write the Burgers equation in terms of the velocity (and gravity) 
potential and we add a noise term, we obtain the Kardar-Parisi-Zhang equation, 
originally proposed for the description of surface growth.
Its fixed point structure corresponding to white noise, as given by the 
dynamical RG, is well known.\cite{Stan} An even richer structure is 
obtained by using power-law correlated noise.\cite{all}


\subsection{Fokker-Planck equation for the Kardar-Parisi-Zhang equation.}

The KPZ equation in Fourier space reads 
\be
{d\phi_{\bf k} \over dt} = F_{\bf k}(\phi_{\bf q}) + \eta_{\bf k}
   = -\nu k^2  \phi_{\bf k} + {\lambda \over 2}
  \sum_{\bf q} {\bm k} \cdot ({\bm{k-q}}) \, 
\phi_{\bf q}\, \phi_{\bf k-q} + \eta_{\bf k}.\label{KPZ}
\ee
It is just a set of coupled Langevin Eqs.\ If the ``force'' $F$ were
linear, then the equations would decouple and actually reduce to
copies of the Langevin Eq.\ for a Brownian oscillator, which is
trivial to solve. 

Let us assume Gaussian correlated noise with 
$\langle \eta_{\bf k}(t)\,\eta_{\bf k'}(t')\rangle = 
                        D(k)\,\delta_{{\bf k}, -{\bf k}'}\,\delta(t-t')$. 
The Fokker-Planck equation for the probability distribution 
$\Pi(\phi_{\bf k},t)$ associated to Eqs.~(\ref{KPZ}) is
\be
{\partial\Pi \over\partial t} = 
-{\partial \over\partial \phi_{\bf k}}[F_{\bf k}\,\Pi] + {1\over 2} 
D(k)\,{\partial^2 \Pi\over\partial \phi_{\bf k} \partial \phi_{\bf -k}}.
\ee
We can compare it with the exact renormalization group equation, under
the assumption of dynamical scaling and, hence, of equivalence of time
and scale evolution, that is, $t \leftrightarrow \ln L$ and
$\Pi(\phi_{\bf k},t) \leftrightarrow \P_L[\varrho]$.  We see that both
equations are approximately equivalent, if $D(k) \leftrightarrow 
L \,(dP_L/dL)$ and $F_{\bf k}(\phi_{\bf q})$ is linear. The former 
condition is interesting, for it relates the freedom in the choice of 
noise with the freedom in the choice of window function, relating 
the fluctuations with the coarse-graining operation. The 
latter condition shows that the nonlinearity tends to suppress an 
exact equivalence. 

\section{Critical phenomena and fractals}

The traditional theory of critical phenomena, including critical
dynamics, applies to situations close to thermodynamical equilibrium,
in which the fluctuations are small. In contrast,
here we are dealing with far-from-equilibrium processes that lead to
very large fluctuations. In fact, the asymptotic state is probably of
fractal nature. In cosmology, a fractal structure of galaxies is
currently discernible on small scales. The scale invariance of fractal
structures has led to assimilate fractal fluctuations with critical
fluctuations and, therefore, the scale at which they subside 
(homogeneity scale) with the correlation length. This is wrong.\cite{us}

Actually, one can obtain for a particle distribution 
with two-point correlation 
$(r_0/r)^{-2\alpha}$ that the relative fluctuation 
of the coarse-grained density is given by 
\be
{{\langle (\delta \rho_{L})^2 \rangle} \over
            \bar \rho^2} = {a^3 \over L^3} + B \left({r_0 \over
              L}\right)^{-2\alpha},
\ee
where $a^3$ is the volume per particle, and $B$ a number.
We clearly see that, for the fractal state,
$L \ll r_0 \Rightarrow \delta\varrho_L \gg {\bar\varrho}$, so that 
thermodynamics is not applicable. Only when $L \gg r_0$ thermodynamics 
becomes applicable as the distribution becomes critical. 

When the fluctuations are very large and actually drive the dynamics,
the use of perturbation theory is questionable. In this sense, it is
also questionable the application of the perturbative dynamical RG to
the gravitational interaction. One should instead employ a
non-perturbative method, as the exact renormalization group, 
advocated here.\footnote{In the more general field of 
dynamical critical phenomena, other non-perturbative approaches
have been proposed.\cite{Pie}}

\section*{Acknowledgments}
I am grateful to A.~Dom{\'\i}nguez, J.~P\'erez-Mercader and 
L.~Pietronero for conversations and to D.~Hochberg for a critical reading 
of this paper.

\section*{Appendix: Proof of the exact renormalization group equation}

There are several proofs of the exact renormalization group 
equation. Here, we adapt the proof provided by T.~Morris\cite{Mor} to
our context.

Let us consider the generator of correlations (characteristic function)
\be
Z[J] := \int\D \varrho\, \P[\varrho]\,\exp(- J\cdot\varrho) =
\int\D \varrho\, \exp\{-{1\over 2}\,{\varrho_L- {\bar \varrho}\over
W_L}\cdot G^{-1}\cdot {\varrho_L- {\bar \varrho}\over W_L} -
V[\varrho_L] - J\cdot \varrho\}.
\ee
As we increase $L \ra L + dL$ the
dynamical variable $\varrho_L \ra \varrho_{L + dL}$, losing detail;
that is, losing high wavenumber modes in the Fourier space. Let us
define the density $\varrho_{dL} := \varrho_{L + dL} - \varrho_{L}$, 
corresponding to the removed high wavenumber modes. In the Fourier space, 
the filtering convolution becomes a multiplication: 
$\varrho_L({\bm k}) =  W_L({\bm k})\,\varrho({\bm k})$. Furthermore, 
we define the filtered power spectrum\cite{Pee}
$P_L({\bm k}) = G({\bm k})\,W_L({\bm k})^2$. Hence, 
$P_{L+dL} = P_L + P_{dL}$. 

We decompose the integral for the generator at the scale $L + dL$ into 
two integrals:
\be
Z[J] = \int\D \varrho_{L}\D \varrho_{dL}\, \exp\{- \sum_{\bf
k}\left[{\varrho_{L}^2\over 2 P_L} +{\varrho_{dL}^2\over 2
P_{dL}}\right] - V[\varrho_{L} + \varrho_{dL}] - J\cdot (\varrho_{L} +
\varrho_{dL})\}. 
\ee
Integrating over $\varrho_{dL}$ to get the generator of correlations of
$\varrho_{L}$, 
\bea
Z_L[\varrho_{L},J] =
\int\D \varrho_{dL}\, \exp\{-{\varrho_{dL}^2\over 2 P_{dL}} -
V[\varrho_{L} + \varrho_{dL}] - J\cdot (\varrho_{L} +
\varrho_{dL})\}\\
= \exp\{{1\over 2} P_{dL}\,J^2- V_L[-P_{dL}\,J+\varrho_{L}] -
J\cdot\varrho_{L}\}\label{VL}
\eea
for a new potential $V_L$.  (To prove it, use the change of variable
$\varrho_{dL} = \varrho_{dL+L} - \varrho_{L}$.)
To unravel
the meaning of $ V_L$, take $J({\bm k}) = 0$ for the modes to be removed 
 ($k>1/L$). Then
$Z_L[\varrho_{L},J] = \exp\{- V_L[\varrho_{L}]-J\cdot\varrho_{L}\},$
showing that it is an {\em effective potential}.

>From its definition,
$${d\over dL}Z_L[\varrho_{L},J] = -{1\over 2}\,{dP_{dL}^{-1}\over dL} 
\left[{\partial \over\partial J({\bm k})}+ \varrho_{L}\right]^2 Z_L$$
Substituting for $Z_L$ and operating,
\be
{dV_L\over dL} = {1\over 2}\,{dP_L\over dL}\left[\left({\partial V_L
\over\partial \varrho_{L}}\right)^2 - {\partial^2 V_L \over\partial
\varrho_{L}^2}\right]= {1\over 2}\,{dP_L\over dL}\,e^{V_L} {\partial^2
\over\partial \varrho_{L}^2} e^{-V_L}.
\ee
The first equation is the usual form of the exact renormalization group 
equation, but the second equality allows us to write it as 
\be
{d\over dL}e^{-V_L} = -{1\over 2}\,{dP_L\over dL}\, {\partial^2
\over\partial \varrho_{L}^2} e^{-V_L}.
\ee
Adding the quadratic part to get the full $\P$ and substituting for 
the power spectrum in terms of $W$, we obtain Eq.~(\ref{ERGeq}).

\end{document}